\documentclass[]{article}
\usepackage{latexsym}
\usepackage{amsmath}
\usepackage{amssymb}
\usepackage{amsfonts}
\usepackage{graphicx}
\usepackage{epsfig}
\usepackage{array}

\newtheorem{ExampleDef}{Example}[section]

\begin{document}
\begin{center}
{\Large {\bf Fast Computation with Neural Oscillators \par}}
\vspace{1.0em}
{\large Wei Wang and Jean-Jacques E. Slotine\footnote{To whom correspondence 
should be addressed.} \par} 
{Nonlinear Systems Laboratory \\
Massachusetts Institute of Technology \\
Cambridge, Massachusetts, 02139, USA 
\\ wangwei@mit.edu, \ jjs@mit.edu 
\par}
\vspace{2em}
\end{center}

\begin{abstract}
Artificial spike-based computation, inspired by models of computations
in the central nervous system, may present significant performance
advantages over traditional methods for specific types of large scale
problems. In this paper, we study new models for two common instances
of such computation, winner-take-all and coincidence detection. In
both cases, very fast convergence is achieved independent of initial
conditions, and network complexity is linear in the number of inputs.
\end{abstract}


\section{Introduction} \label{sec:introduction}

Recent research has explored the notion that artificial spike-based
computation, inspired by models of computations in the central nervous
system, may present significant advantages for specific types of large
scale problems~\cite{hopfield, spiking_sum, binding, jin02, llinas03, lazzaro}.  
This intuition is motivated in part by the fact
that while neurons in the brain are enormously "slower" than silicon
based elements (about six orders of magnitude in both elementary
computation time and signal transmission speed), their performance in
networks often compares very favorably with their artificial
counterparts even when reaction speed is concerned. In a sense,
evolution may have been forced to develop extremely efficient
computational schemes given available hardware limitations.

In this paper, we study new models for two common instances
of such computation, winner-take-all and coincidence detection. In
both cases, very fast convergence is achieved and network complexity
is linear in the number of inputs. 

We first present a simple network of FitzHugh-Nagumo (FN) neurons for
fast winner-take-all computation. In contrast to most existing
studies, e.g. the recent~\cite{jin02}, the network's initial state
can be arbitrary, and its convergence is guaranteed in at most two
spiking periods, making it particularly suitable to track time-varying
inputs. If several neurons receive the same largest input, they all
spike as a group.

Using a very similar architecture, but replacing global inhibition by
global excitation, we obtain an FN network for fast coincidence detection,
in a spirit similar to~\cite{hopfield}. Again the system's
response is practically immediate, regardless of the number of inputs.

In section 2 we review basic properties of the FitzHugh-Nagumo model
in its dimensionless version. Sections 3 and 4 discuss applications to
the design of fast winner-take-all networks and coincidence detection
networks. Brief concluding remarks are offered in Section 5.

\section{The FitzHugh-Nagumo Model} \label{sec:f-n}

The FitzHugh-Nagumo model~\cite{fitzhugh-nagumo} is a well-known
simplified version of the classical Hodgkin-Huxley
model~\cite{hodgkin-huxley}, the first mathematical model of wave
propagation in squid nerve. Originally derived from the Van der Pol
oscillator~\cite{books_vdp}, it can be generalized using
a linear state transformation to the dimensionless
system~\cite{murray}
\begin{equation} \label{eq:f-n}
\begin{cases}  
  \ \dot{v} = v(\alpha - v)(v-1)-w+I  \\  
  \ \dot{w} = \beta v - \gamma w
\end{cases}  
\end{equation}
where $\alpha, \beta, \gamma$ are positive constants. Here $v$ models membrane
potential, $w$ accommodation and refractoriness, and $I$ stimulating
current.

Simple properties of the FN model~\cite{murray} can be exploited for
neural computations. For appropriate parameter choices, there exists
a unique equilibrium point for any given value of $I$. Furthermore,
this equilibrium point is stable, except for a finite range $\ I_l \le
I \le I_h\ $ where the system tends to a limit cycle.  The
steady-state value of $v$ at the stable equilibrium point increases
with $I$.


%
\section{Winner-Take-All Network} \label{sec:wta}
Winner-take-all (WTA) networks, which pick the largest element from a
collection of inputs, are ubiquitous in models of neural computation,
and have been used extensively in the contexts of competitive
learning, pattern recognition, selective visual attention, and
decision making \cite{wta_sum, deliang99, yuille02}. Furthermore,
Maass~\cite{maass00} showed that WTA represents a powerful computational
primitive as compared to standard neural network models based on
threshold or sigmoidal gates.

The architectures of most existing WTA models are based on inhibitory
interactive networks, implemented either by a global inhibitory unit
or by mutual inhibitory connections. Many studies, such
as~\cite{old_wta}, require the system dynamics to be
initiated from a particular state, which prevents real-time tracking
of time-varying inputs. Starting with~\cite{lazzaro}, many
WTA implementations in analog VLSI circuits have been
proposed. While they do guarantee a unique global minimum, dynamic
analysis is difficult and computation resolution limited. Studies of
spike-based WTA computation, as in~\cite{jin02}, are comparatively
recent.  In this section, we describe a very simple network of
FitzHugh-Nagumo neurons for fast winner-take-all computation, whose
complexity is linear in the number of inputs. The network's initial
state can be arbitrary, and its convergence is guaranteed in at most
two spiking periods, making it particularly suitable to track
time-varying inputs. If several neurons receive the same largest
input, they all spike as a group of winners.

\begin{figure}[h]
\begin{center}
\epsfig{figure=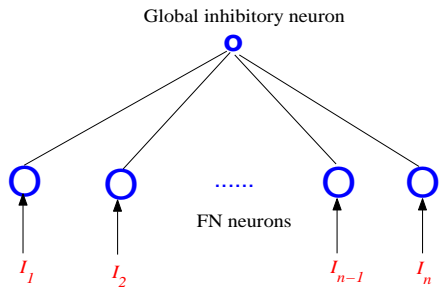,height=40mm,width=60mm}
\caption{Diagram of the network. There are $n$ FN neurons receiving 
different external stimulating inputs. A global inhibition neuron
monitors the network.}
\label{fig:structure}
\end{center}
\end{figure}
As illustrated in Figure~\ref{fig:structure}, the network consists of 
$n$ FN neurons. Each neuron receives a stimulating input $I_i$ from outside 
as well as a common inhibition current $z$ from the global inhibition 
neuron. The dynamics of the FN neurons ($i=1,\ldots,n$) are
\begin{equation*} \label{eq:fn-in-wta}
\begin{cases}  
  \ \dot{v}_i = v_i (\alpha - v_i) (v_i-1)-w_i + I_i - z  \\  
  \ \dot{w}_i = \beta v_i - \gamma w_i
\end{cases}  
\end{equation*}
The dynamics of the global inhibition neuron switches between a
charging mode and a discharging mode. It starts charging if there is
any FN neuron spiking in the network, i.e. when $v_i$ exceeds a
given threshold value $v_0$.  It switches to discharging if the state
is saturated (enough close to the saturation value in simulation)
and stays at this mode until next time a FN neuron
spikes. The specific dynamics of these two modes can be very
general. For simplicity, we use
\begin{equation*} \label{eq:inhibition-in-wta}
\dot{z} = 
\begin{cases}  
  \  - k_c \ (z - z_0)  \ \ \ \ \mathrm{charging}\ \mathrm{mode} \\  
  \  - k_d \ z   \ \ \ \ \ \ \ \ \ \ \ \ \mathrm{discharging}\ \mathrm{mode}
\end{cases}  
\end{equation*}
where $z_0$ is a constant saturation value, and $k_c$ and $k_d$ are the
charging rate and discharging rate.

Within at most two periods, the winner will be the only neuron
spiking in the whole network. In some cases, this is achieved
within the first period. 

\begin{figure}[h]
\begin{center}
\epsfig{figure=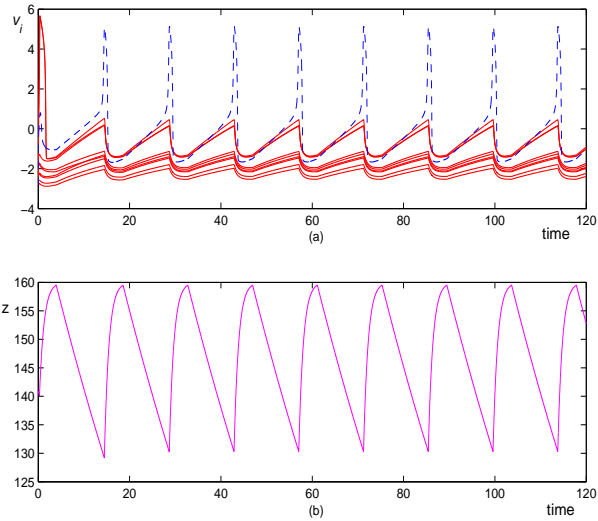,height=70mm,width=80mm}
\caption{ Simulation result of WTA computation with $n=10$.  The
parameters of the FN neurons are set as $\alpha = 5.32, \beta = 3,
\gamma = 0.1$, with spiking threshold $\ v_0 = 5\ $.  The inputs $I_i$
are chosen randomly from $20$ to $125$. The parameters of the global
neuron are $z_0 = 160, k_c = 1, k_d = 1/50$. All the initial
conditions are chosen arbitrarily. (a) States $v_i$ versus time. The
dashed curve represents the state of the neuron receiving the largest
input. (b) State $z$ versus time.}
\label{fig:wta-normal}
\end{center}
\end{figure}

To perform WTA computation, we set the charging rate of the global
neuron to be fast and the discharging rate to be slow. Thus, similarly
to~\cite{jin02}, if there is any FN neuron spiking, the strength of
the inhibition current increases to its saturation value very rapidly,
leaving no chance for the other neurons to spike. The global neuron
then discharges slowly, which lets the FN neurons smoothly approach
the oscillation region. The first neuron entering the oscillation
region will be the one with the largest input. So it spikes as the
winner and ignites a new period. Note that before enter the
oscillation region, all FN neurons converge to their equilibrium
points with the equilibrium point of the winner having the largest
value. A slowly discharging process allows the winner to occupy the
highest position and helps it to spike immediately once it enters the
oscillation region.  Given the parameters of the FN neurons, the
frequency of the result depends on the global neuron's saturation
value, its charging and discharging rates, and the value of the
largest input. If we also fix the global neuron dynamics, the
frequency increases with the increasing of the largest input.
Simulation results are shown in Figure~\ref{fig:wta-normal}.

\noindent {\em {\large Remarks}}

\noindent $\bullet$ {\bf Initial conditions and computation speed}\ \
\ \ \ The mechanism described above guarantees that initial conditions
can be set arbitrarily, which cannot be realized by most of the
previous WTA models. With appropriate parameters, the computation can
be completed at most in two periods. The first spiking neuron is
chosen by initial conditions, while the second one is the neuron with
the largest input, which remains the winner until the inputs change.
Actually, if the initial inhibition is set large enough so that all
the FN neurons are depressed in the beginning, then the neuron with
the largest input will spike first. The computation speed of our FN
network is faster and more robust than the WTA model recently
presented in~\cite{jin02}, whose network of integrate-and-fire neurons
has to wait until the winner gets the right to spike, which may take a
long time for large networks.

\begin{figure}[h]
\begin{center}
\epsfig{figure=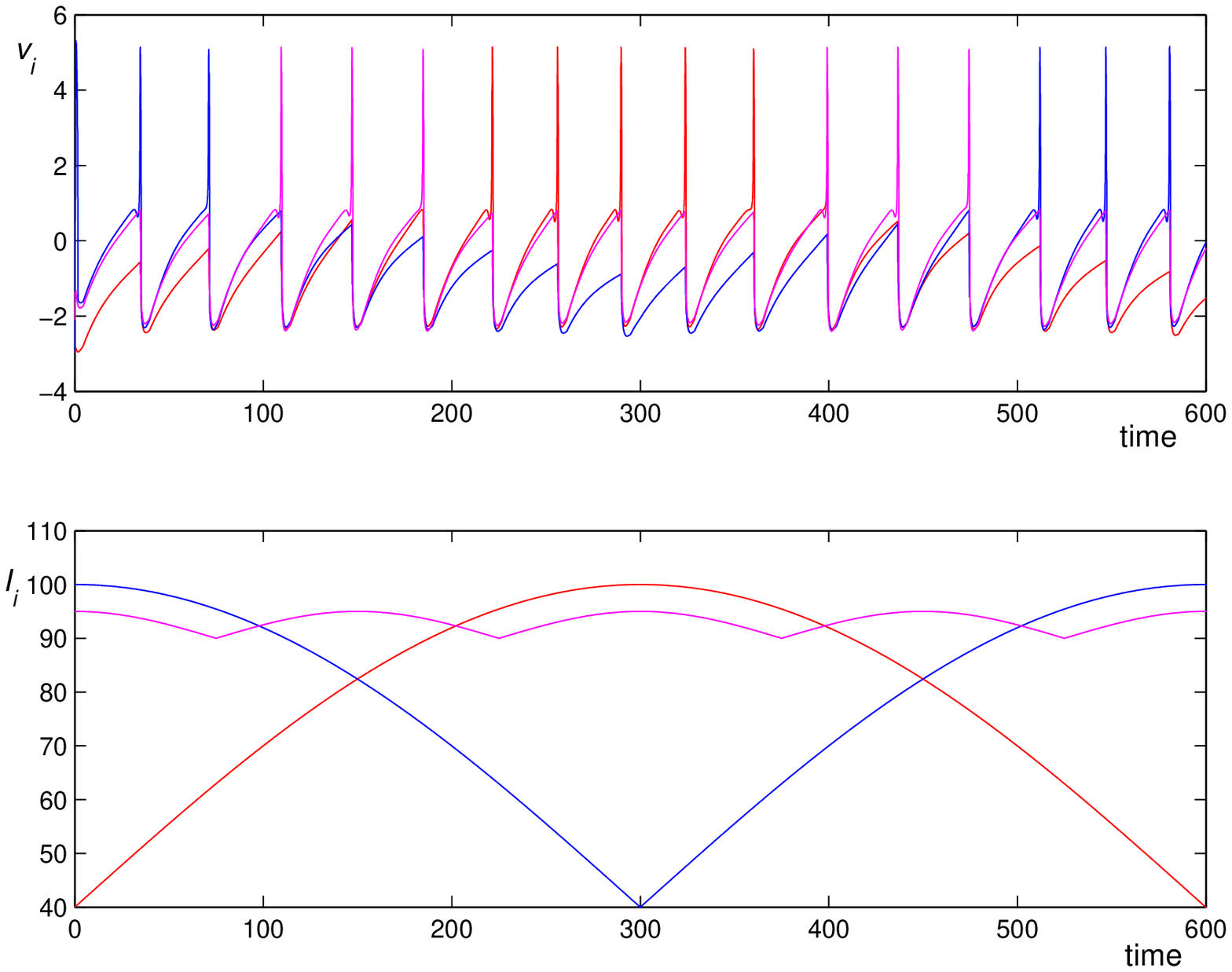,height=70mm,width=80mm}
\caption{ Simulation result of WTA computation with varying input. The
parameters are all equal to those in Figure~\ref{fig:wta-normal}
except that $n=3$. Inputs change continuously. The winner always
tracks the neuron with the largest input.}
\label{fig:wta-varying}
\end{center}
\end{figure}

\noindent $\bullet$ {\bf Varying inputs and noise} \ \ \ \ \ Since
initial conditions do not matter in our model, the network can easily
track time-varying inputs. Figure~\ref{fig:wta-varying} illustrates
such an example, where three inputs switch winning positions several
times. The spiking neuron always tracks the largest input. The
computation is robust to signal noise as well.

\begin{figure}[h]
\begin{center}
\epsfig{figure=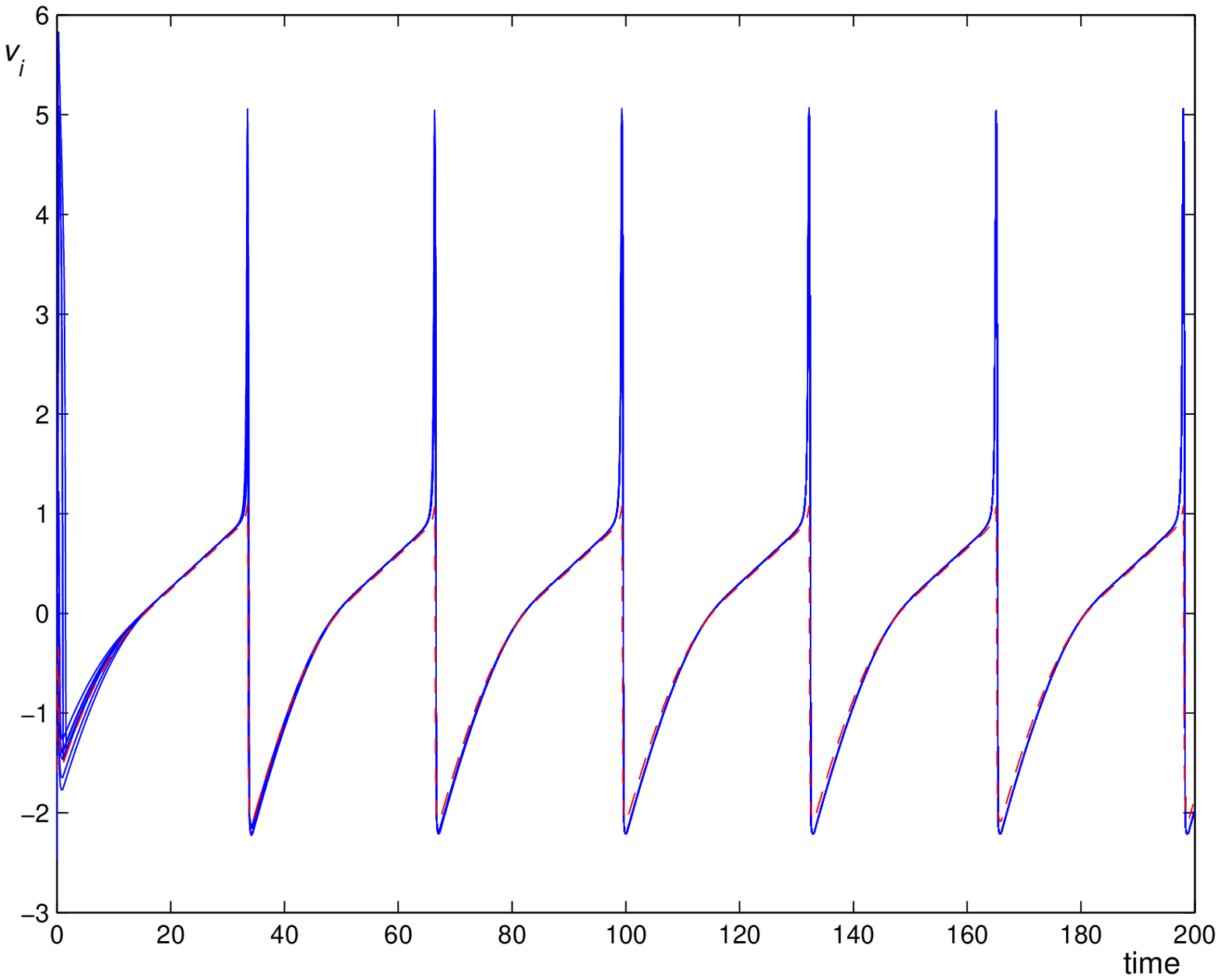,height=40mm,width=70mm}
\ \ \ \ \ \ \ \ \epsfig{figure=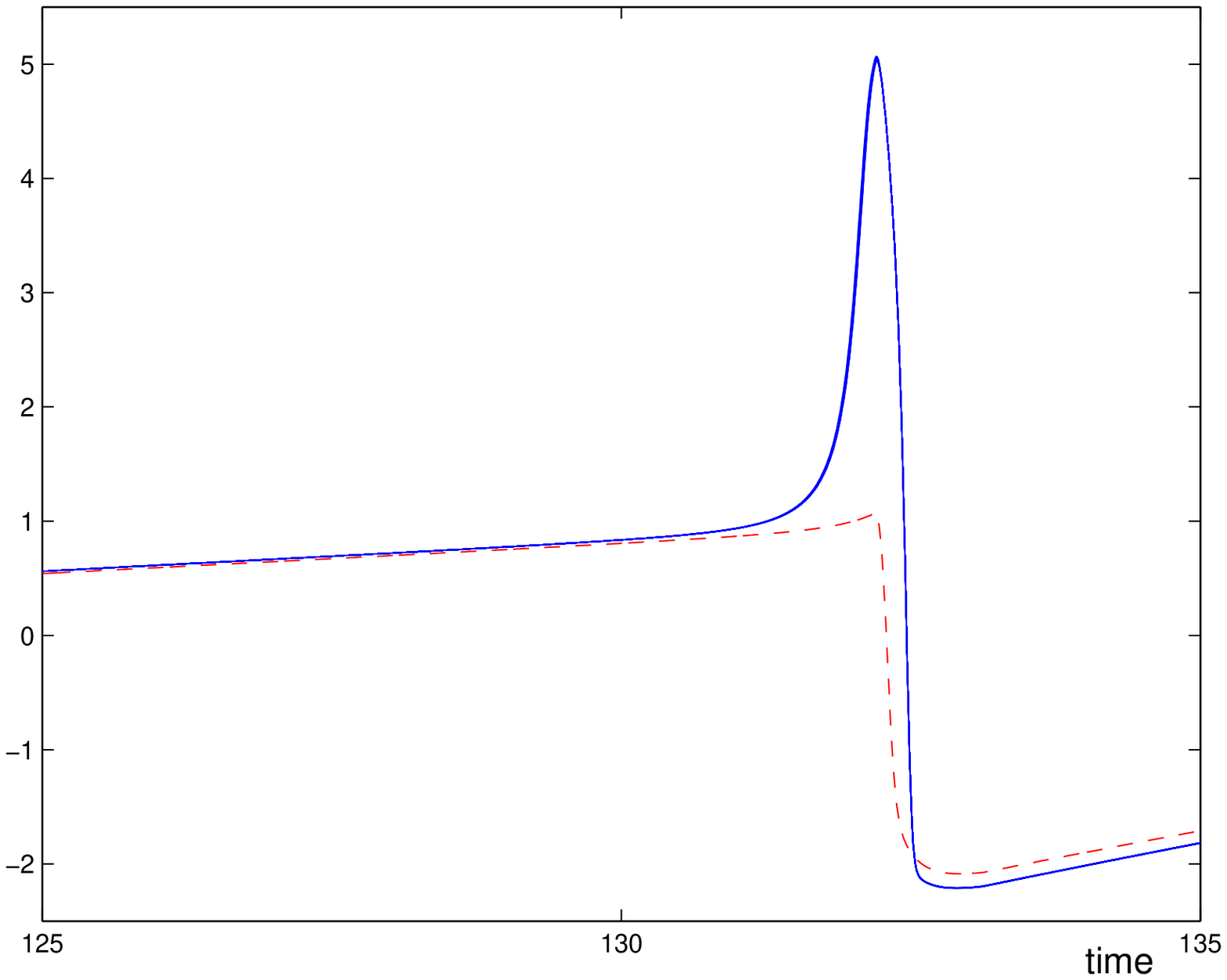,height=40mm,width=20mm}
\caption{Simulation result of WTA computation with multiple
winners. The parameters are the same as in Figure~\ref{fig:wta-normal}
except that $k_c = 5,\ k_d = 1/80$.  The inputs are $\ I_1 = \cdots =
I_9 = 120, I_{10} = 119.5\ $. The first plot shows the $v_i$'s as
functions of time. The dashed (red) line represents $v_{10}$ and the
solid (blue) lines the other $v_i$'s. The first nine neurons converge together
during the waiting time and spike simultaneously as a group of
winners. The second plot, an enlarged version of the first at a
spiking moment, shows that $v_{10}$ is completely depressed by the
winners even though the input difference is very small.}
\label{fig:wta-multiple}
\end{center}
\end{figure}

\noindent $\bullet$ {\bf Multiple winners} \ \ \ \ Decreasing the
global neuron discharging rate $k_d$ extends the waiting time before
the winner enters the oscillation region. This is helpful if there
exist several neurons receiving the same largest input and we expect
them all spike as a group of winners. Enlarging the time neurons stay
in the stable region allows these neurons with the same input converge
to each other, and to enter the oscillation region and spike
simultaneously ( Figure~\ref{fig:wta-multiple}).  Note that
in~\cite{jin02}, only one winner can succeed and it is picked
arbitrarily from the group of candidates.

If the network size is small, the network may be augmented with
all-to-all couplings {\it between} FN neurons, with the coupling gain
increasing with the similarity of the inputs (e.g., of the form
$e^{- \ \alpha |\ I_j - I_i \ |}$ ). This lets the neurons
receiving identical inputs converge together exponentially (using
partial contraction theory \cite{wei}) and thus provides
another solution to the multiple-winner problem.

\noindent $\bullet$ {\bf Computation resolution} \ \ \ \ Computation
resolution can be improved by decreasing the global neuron discharging
rate $k_d$ while increasing the charging rate $k_c$. Decreasing $k_d$
allows the winner fully distinguished with the following neurons;
increasing $k_c$ prevents the following neurons spike after the
winner. Figure~\ref{fig:wta-multiple} illustrates such an example with
winners $I_{max} = 120$ while the second largest input $I_{10} =
119.5$. The resolution here is much better than the WTA models
presented previously, including~\cite{lazzaro,urahama}. It can be
further enhanced by decreasing the relaxation time of the FN neurons.

\noindent $\bullet$ {\bf Input bounds} \ \ \ \ The inputs to the FN
neurons should be lower-bounded by $I_l$ (the lower threshold of the
oscillation region) to guarantee that the neurons can spike before the
inhibition is fully released. They should also be upper bounded to set
$z_0$.

\noindent $\bullet$ {\bf $K$-Winner-Take-All} \ \ \ \
$K$-Winner-Take-All ($k$-WTA) is a common variation of WTA
computation~\cite{maass00}, where the output indicates for
each neuron whether its input is among the $k$ largest.  An example
$k$-WTA circuit in~\cite{urahama} extends the WTA model
in~\cite{lazzaro} by formulating the problem in terms of mathematical
programming, but it inherits its low resolution limit
from~\cite{lazzaro} as well. Conversely, the advantages of the above
FN network generalize to the $k$-WTA case.  As inhibition decreases,
the FN neurons enter the oscillation region rank-ordered by their
inputs. For WTA computation, we charge the inhibition neuron after the
first arrival. For $k$-WTA computation, we only need to modify the
charging moment to capture the $k^{\rm th}$ arrival instead. Since the
neurons enter the oscillation region in sequence, they spike in
sequence. If we set $k=n$, we get a pre-ordered spiking sequence in
each period, which may be used to realize soft-WTA~\cite{maass00,
yuille02}, and also provides a simple desynchronization mechanism for
binding problems~\cite{binding}. The
computation resolution follows directly from that in WTA.  A detailed
description and discussion of the $k$-WTA network will be presented
separately.

\noindent $\bullet$ {\bf Spike-controlled coupling and slow
inhibition} \ \ \ \ The feedforward and recurrent connections used in
our WTA network are similar to those in~\cite{llinas03}, where a
``universal'' control system is developed based on olivo-cerebellar
networks. The couplings inside the circuit are also spike-controlled
and they use a FitzHugh-Nagumo-like model containing four variables. A
similar mechanism is also used in~\cite{deliang99}, where WTA is
implemented to compute the object with the largest
size. Slowly-discharged inhibition is also used in biologically
motivated models such as~\cite{slow_inhinition, deliang99}.

\noindent $\bullet$ {\bf Computational complexity} \ \ \ \ \ 
The complexity of the network is $O(n)$. Since the FN neurons are
independent, they can be added or removed from the network at any
time.

\section{Fast Coincidence Detection} \label{sec:fsd}
Recent neuroscience research suggests that coincidence detection plays
a key role in temporal binding~\cite{coincidence}. Hopfield {\em
et al.}~\cite{hopfield} proposed two neural network
structures, both able to capture a ``many-are-equal'' moment, to model
speech recognition and olfactory processing. A similar computation can
be implemented by FN neurons, with faster and more salient response.

\begin{figure}[h]
\begin{center}
\epsfig{figure=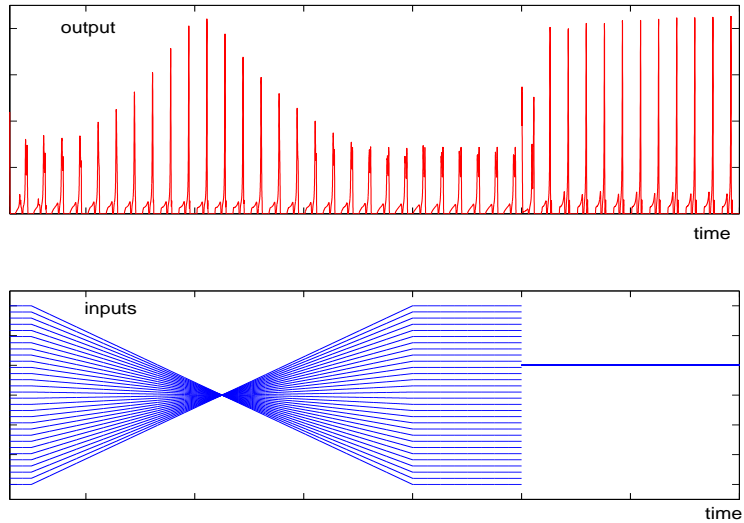,height=70mm,width=100mm}
\caption{Simulation result of fast coincidence detection with
$n=30$. The leader and the followers are FN neurons with the same
parameters as in Figure~\ref{fig:wta-normal}. The inputs are $I_0=90$
and $I_1, \ldots, I_{30}$ varying from $20$ to $80$, and the coupling
gain is $k=1.7$. The upper plot shows $\ \sum_{i=1}^{n} \max (0,
\dot{v_i})$ versus time, and the lower $I_1, \ldots, I_n$.}
\label{fig:hopfield}
\end{center}
\end{figure}

Consider a leader-followers network with a structure similar to
Figure~\ref{fig:structure}, except that the global neuron (the leader)
is now {\it excitatory}, and the connections from the leader to the
followers are unidirectional. For simplicity, we assume that all the
neurons are FN neurons with the same parameters but different
inputs. The dynamics of the leader $(v_o , w_o)$ obeys
equations~(\ref{eq:f-n}) while those of the followers ($i=1,\ldots,n$)
are
\begin{equation*}
\begin{cases}  
  \ \dot{v}_i = v_i(\alpha - v_i)(v_i-1)-w_i+I_i + k(v_0-v_i)  \\  
  \ \dot{w}_i = \beta v_i - \gamma w_i
\end{cases}  
\end{equation*}
where $k(v_0-v_i)$ is the coupling force from the leader to the
followers.  Neurons $i$ and $j$ synchronize only if inputs $I_i$ and
$I_j$ are identical.  We define the system output accordingly to
capture the moment when this condition becomes true for a large number
of inputs, as illustrated in Figure~\ref{fig:hopfield}. Note that the
coupling gain $k$ should be large enough to guarantee synchronization
(an explicit threshold can be computed
analytically~\cite{wei}), but not so large as to have the
leader numerically dominate the dynamic differences between the
followers. More general formal studies of synchronization can be found
in~\cite{wei}, based on nonlinear contraction theory.

\section{Concluding Remarks} \label{sec:conclusion}
Basic computations such as winner-take-all and coincidence detection
can be performed fast and robustly using extremely simple spike-based
models. The results are currently being extended to higher-level
perception problems.

\vspace{1.5em}

\noindent {\large{\bf Acknowledgments:}} This work was supported in
part by a grant from the National Institutes of Health. The authors
benefited from stimulating discussions with Matthew Tresch.


%
%
\renewcommand{\baselinestretch}{1.}

\end{document}